%% file: WLSFiberV2.tex
\documentclass[12pt]{article}
\usepackage{graphicx}
\usepackage{authblk}
\usepackage{color}
\usepackage{wasysym}
\usepackage[small,labelfont=bf,up]{caption}

\newcommand\pubnumber{DPF2015-350}
\newcommand\pubdate{\today}

\def\napoli{on behalf of the Mu2e Collaboration}
\def\UVA{University of Virginia}
\def\Shooltz{Shooltz Solutions LLC}

\def\Title#1{\begin{center} {\bf\Large #1 } \end{center}}

\def\Address#1{\begin{center}{\textbf{\it #1}} \end{center}}

\newcommand\pubblock{\rightline{\begin{tabular}[!htbp]{l} \pubnumber\\
         \pubdate  \end{tabular}}}
\newenvironment{Abstract}{\begin{quotation}  }{\end{quotation}}
\newenvironment{Presented}{\begin{quotation}\begin{center}
\hspace{-5mm}PRESENTED AT\end{center}\bigskip
      \begin{center}\begin{large}}{\end{large}\end{center} \end{quotation}}
\def\Acknowledgments{\bigskip  \bigskip \begin{center} \begin{large}
             \bf ACKNOWLEDGMENTS \end{large}\end{center}}

\input{econfmacros.tex}

\title{\Title{Performance of Wavelength-Shifting Fibers for the Mu2e Cosmic Ray Veto Detector}}

\author[$\dag$]{G.~DeZoort}
\author[$\dag$]{E.~C.~Dukes}
\author[$\dag$]{R.~C.~Group}
\author[$\dag$]{H.~Kessenich}
\author[$\dag$]{Y.~Oksuzian}
\author[$\dag$]{T.~Rase}
\author[$\S$]{D.~Shooltz}
\affil[$\dag$]{\UVA}
\affil[$\S$]{\Shooltz}

\date{}

\begin{document}
\begin{titlepage}
\pubblock
\vspace{-5mm}
{\let\newpage\relax\maketitle}
\thispagestyle{empty}
\vspace{-7mm}
\Address{\napoli}
\vspace{7mm}
\begin{Abstract}
The Mu2e experiment will search for a neutrino-less muon-to-electron
conversion process with almost four orders of magnitude of sensitivity
improvement relative to the current best limit. One important
background is caused by cosmic ray muons, and particles produced by
their decay or interactions, mimicking the conversion electron
signature. In order to reach the design sensitivity, Mu2e needs to
obtain a cosmic ray veto (CRV) efficiency of 99.99\%. The CRV system
consists of four layers of plastic scintillating counters read out by silicon
photo-multipliers (SiPM) through wavelength shifting fibers. The CRV
counters must produce sufficient photo statistics in order to achieve
the required veto efficiency. We study the light properties of several
wavelength shifting fiber sizes in order to optimize the total light
yield for the CRV system. The measurements are performed using a
scanner designed to ensure fiber quality for the CRV.
\end{Abstract}
\vspace {5mm}
\begin{Presented}
DPF 2015\\
The Meeting of the American Physical Society\\
Division of Particles and Fields\\
Ann Arbor, Michigan, August 4--8, 2015\\
\end{Presented}
\vfill
\end{titlepage}
\def\thefootnote{\fnsymbol{footnote}}
\setcounter{footnote}{0}

\section{Introduction}
The Mu2e experiment \cite{tdr} proposes a search for neutrino-less
conversion of a muon into an electron in the presence of a
nucleus. In the Standard Model(SM) this process can occur due to
neutrino oscillations but at a rate approximately thirty orders of
magnitude below the current best experimental limits. However, many SM
extensions suggest that this process will occur at an enhanced rate
that would be observable experimentally. An observation of this conversion at Mu2e would be an
unambiguous sign of physics beyond the SM. The total background
budget at Mu2e is less than 0.4 events after three years of data
taking. However, with out a veto system one background event will be produced per
day at Mu2e due to cosmic ray muons. So, in order to achieve the
required sensitivity, Mu2e needs to suppress the cosmic ray background
by four orders of magnitude. The cosmic ray veto (CRV) system has been
designed to cover the Mu2e apparatus and veto cosmic ray background
with 99.99\% efficiency. The CRV consists of plastic scintillator
counters read out through wavelength shifting (WLS) fibers. The CRV
efficiency critically depends on the light yield emission from the WLS
fibers. We have designed and built a WLS fiber scanner to measure the
light properties and ensure the quality of the fiber used in the CRV.

\section{Wavelength shifting fiber}
\begin{figure}[htb]
\centering 
\includegraphics[height=2.3in]{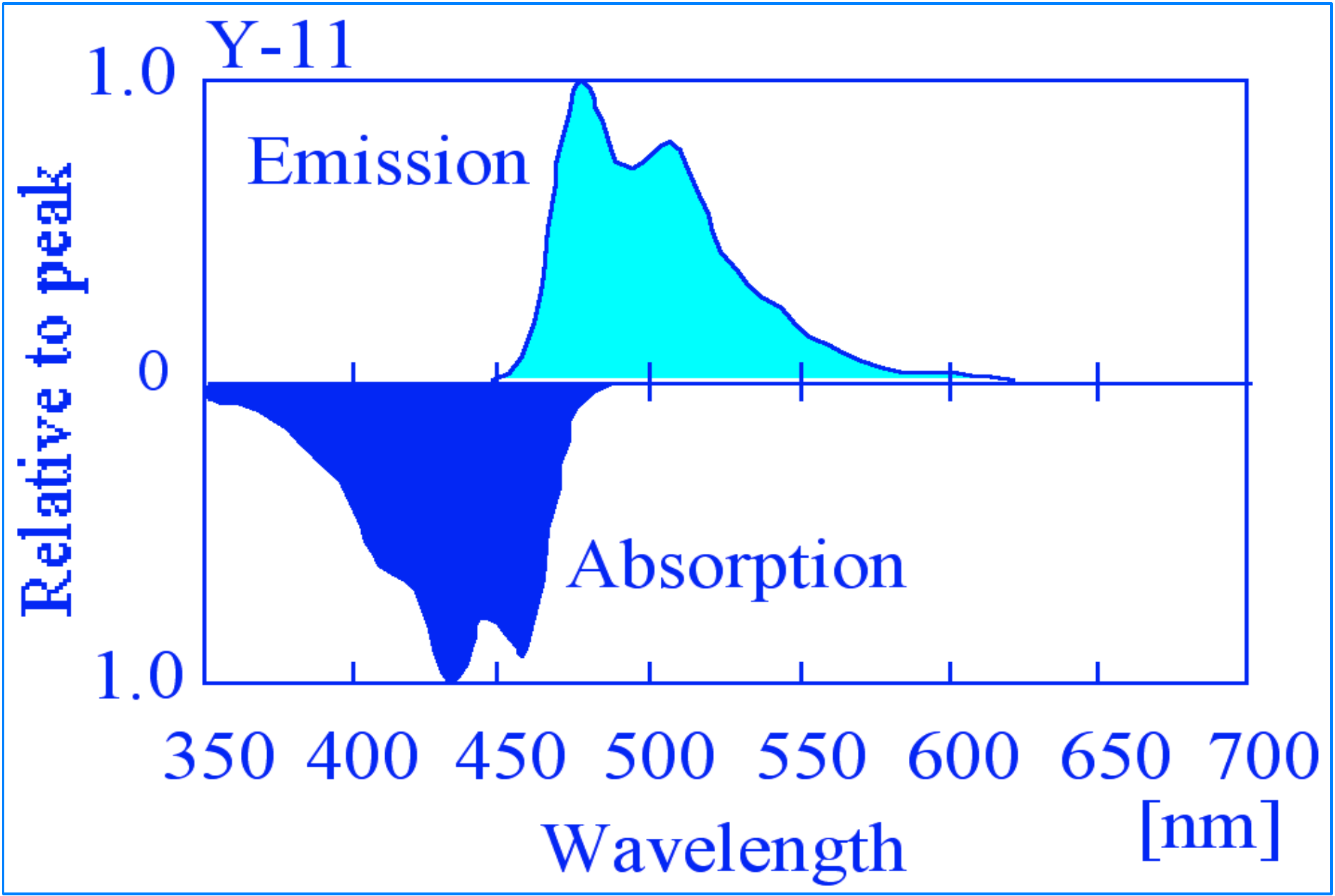}
\includegraphics[height=2.4in]{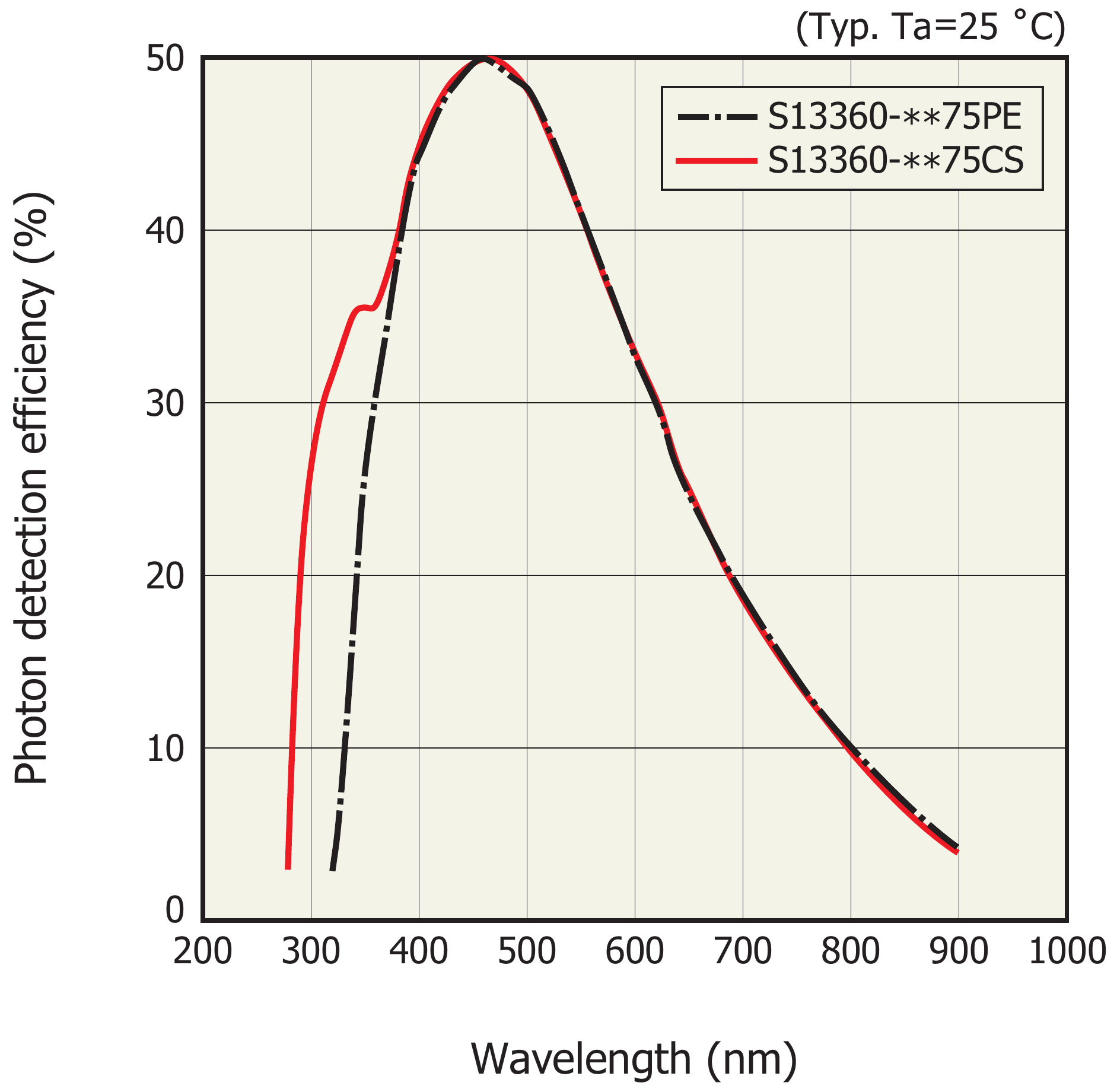}
\caption{The absorption and emission spectra of Kuraray Y11 fiber. 
The spectral PDE response of Hamamatsu SiPM.}
\label{fig:spectra}
\end{figure}

The Mu2e experiment plans to use Kuraray~\cite{kuraray} Y11 WLS fibers
for the CRV system. The fluorescent dye, K27, in the Kuraray Y11
fibers traps the blue light ($\sim$425 nm) from the
scintillating counters and re-emits the light in the green (500-600
nm) spectral region (Figure~\ref{fig:spectra}). The fibers are read
out by silicon photomultipliers (SiPM). The spectral photon
detection efficiency response of the chosen SiPM \cite{sipm} type has
is well match to the emission spectrum of Kuraray Y11 fiber. We have
selected multi-clad and non-S type fiber due to the enhanced light yield
that it provides~\cite{y11}

\section{Fiber scanner}
\begin{figure}[htb]
\centering 
\includegraphics[height=3.0in]{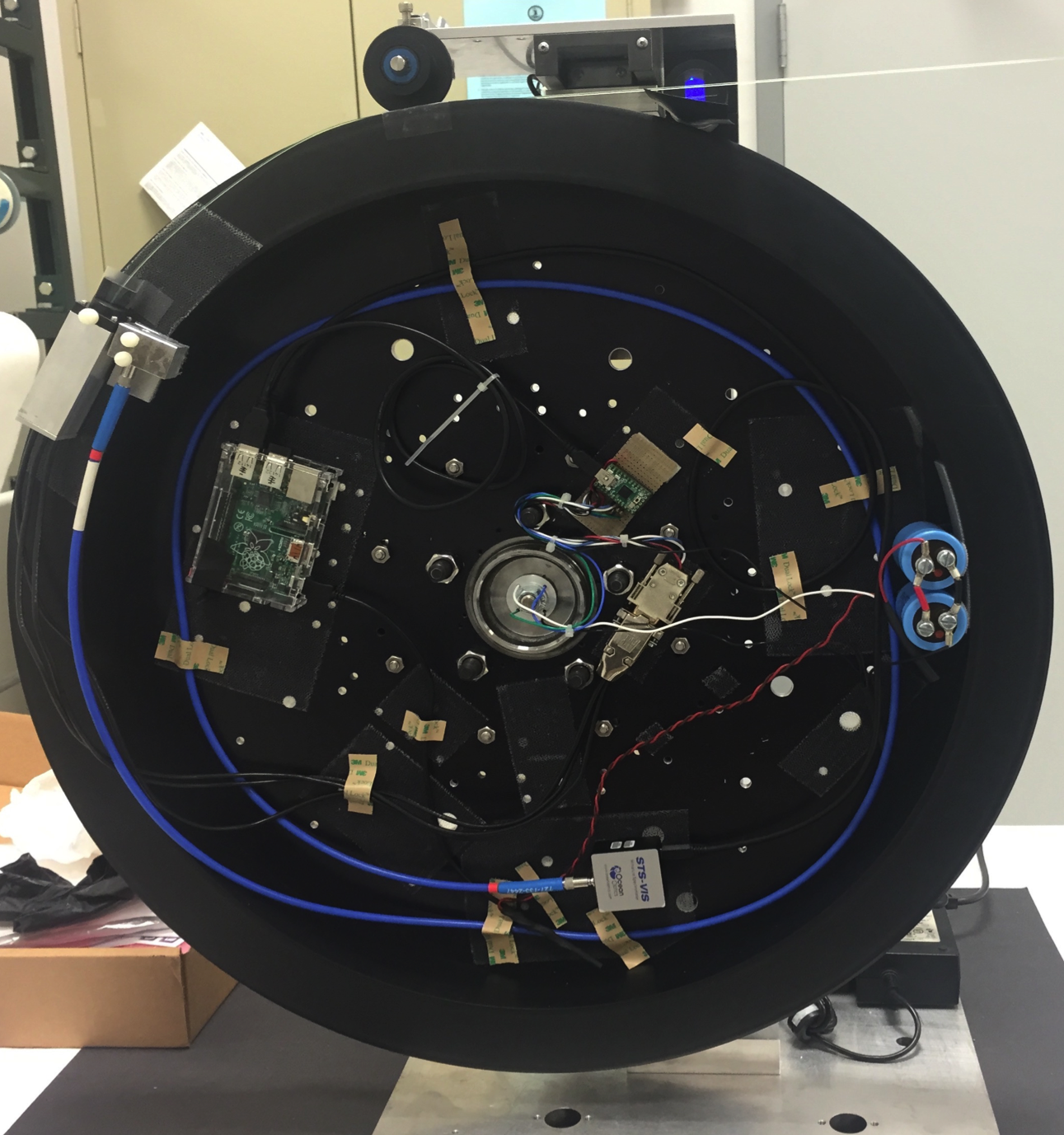}
\caption{Fiber scanner.}
\label{fig:scanner}
\end{figure}

\begin{figure}[htb]
\centering 
\includegraphics[height=3.0in]{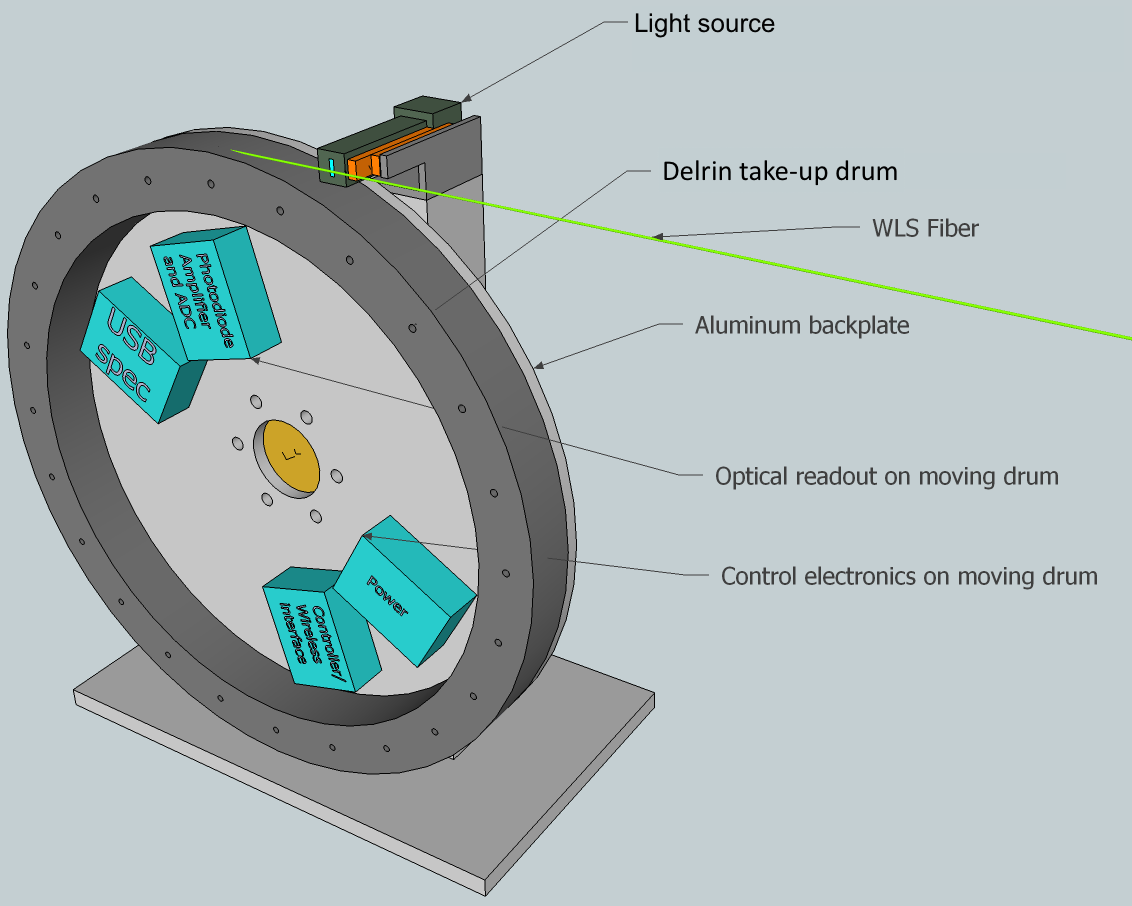}
\caption{Schematics of the fiber scanner.}
\label{fig:schematics}
\end{figure}

We designed and produced a fiber scanner to study the fiber properties
and to assure the quality of the fibers used for the CRV system (see a
photo in schematic drawing in Figures~\ref{fig:scanner,fig:schematics})
. The fiber scanner consists of a large diameter ($\diameter$ = 62 cm)
take-up drum carrying the optical readout hardware. The large diameter
of the take-up drum was selected to accomodate the large-diameter ($\diameter <$
0.2 cm)  WLS fibers that we are considering for the CRV system. The
recommended minimum bending diameter \cite{y11} for this type of fiber
is below the diameter of the take-up drum.

A blue LED light source excites the WLS fiber and is read out by a
large area Hamamatsu S1227-1010BR photodiode \cite{diode} and STS-VIS
USB Ocean Optics spectrophotometer \cite{sts}. The take-up drum is
driven by a stepping motor and controlled for gentle acceleration  and
deceleration. The fiber is delivered by the manufacturer on 90 cm diameter cardboard
spools. The fiber end is connected to one of the readout devices
during the winding process. The take-up drum can increment at any
predetermined distance, stopping for measurements of the light
spectra and intensity. A fiber scan procedure is performed remotely
through a web interface on a Raspberry Pi mounted on the take-up
drum. During data taking the Raspberry Pi controls the step motor,
and collects and stores the data from the photodiode and spectrometer.

\section{Results}
\begin{figure}[htb]
\centering
\includegraphics[height=3.0in]{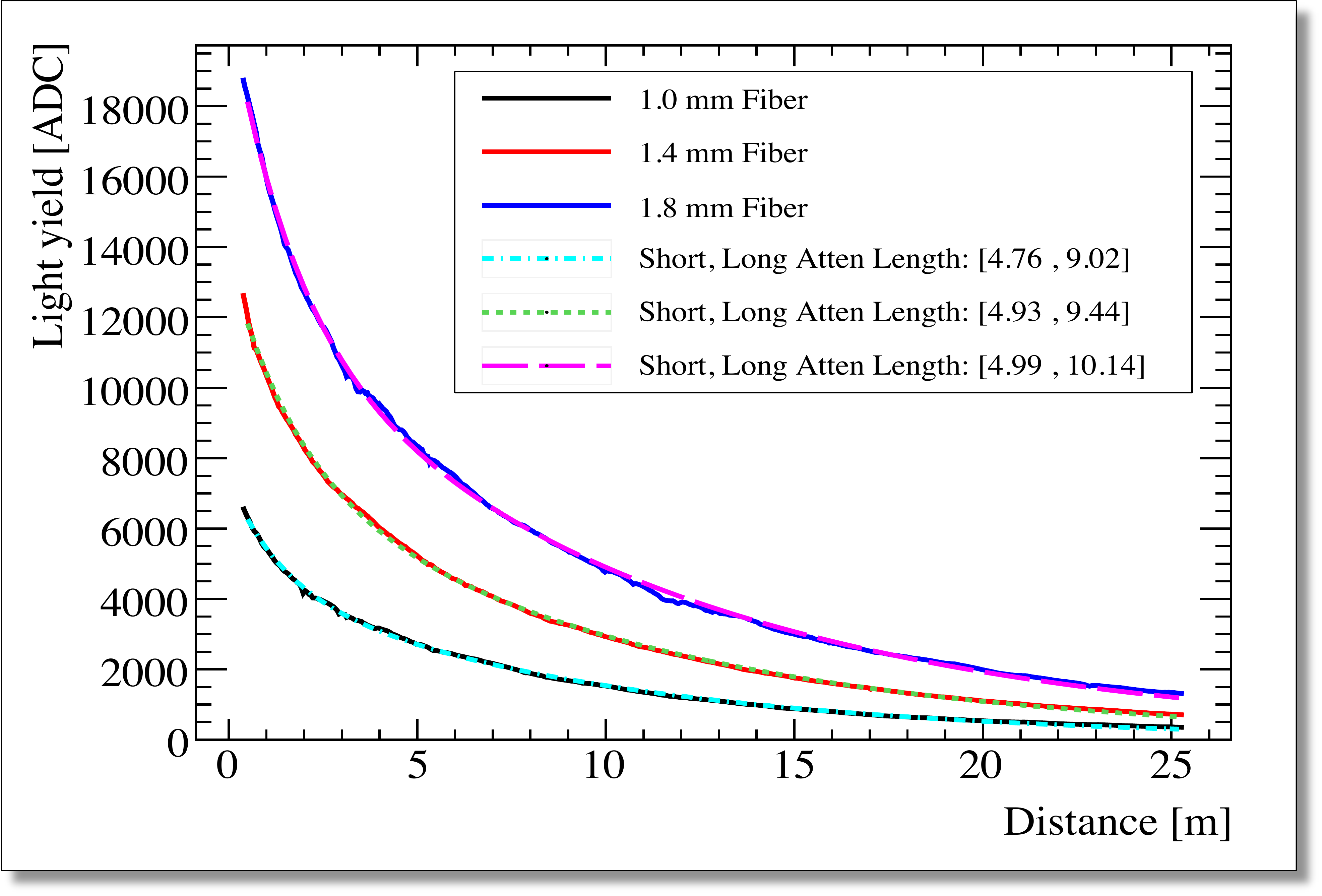}
\caption{Light yield measurements for 1.0, 1.4 and 1.8 mm Kuraray Y11
  WLS fiber, using fiber scanner and photodiode readout scheme.}
\label{fig:diodeatten}
\end{figure}

The photodiode measures the absolute value of the light yield. The
large photodiode area provides stable measurements, which are
not sensitive to the effects from a fiber misalignment or
orientation. The photodiode yields a uniform response \cite{diode} in
the wavelength spectral region of the WLS fiber emission. The spectral response of
the SiPM used in the CRV system features a peak sensitivity at 450
nm (Figure~\ref{fig:spectra}). Therefore, the light attenuation
measured in the CRV system slightly differs from the one measured with the
photodiode. Even though the light attenuation measurements using
the photodiode and SiPM can not be directly compared, a photodiode scan
can identify compromised fiber with sharp drops in
the light yield or poor attenuation. The result of the photodiode scan
is presented in Figure~\ref{fig:diodeatten} and shows the light yield
measurements at 5 cm increments over 25 m of fiber. The short
and long attenuation components are extracted from the fit to a double
exponential decay function: $f(x) =
Ae^{-x/{\lambda_S}}+Be^{-x/{\lambda_L}}$. We have examined 1.0, 1.4
and 1.8 mm diameter fibers. We have observed that the light yield for
1.4 and 1.8 mm fiber is higher than the light yield from 1.0 mm fiber
by factors of 2 and 3 respectively. In addition to better light yield,
thicker diameter fibers provide slightly better attenuation as shown
in Figure~\ref{fig:diodeatten}.

\begin{figure}[htb]
\centering
\includegraphics[height=3.0in]{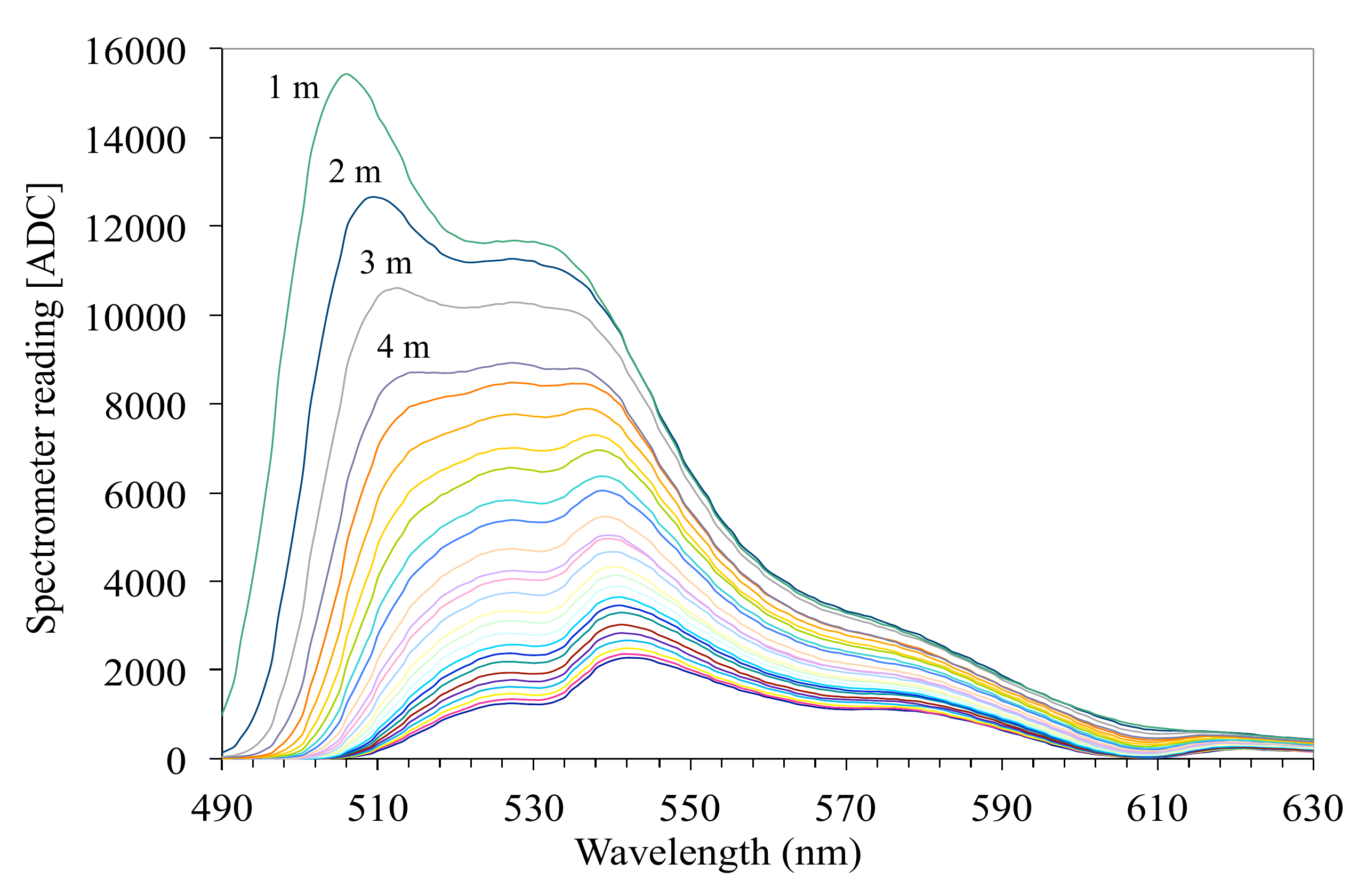}
\caption{Spectrometer readings as a function of fiber emission
  wavelength. Individual curves correspond to 
  measurements taken at various distances between fiber excitation
  points and the spectrometer readout.}
\label{fig:spectromreads}
\end{figure}

\begin{figure}[htb]
\centering
\includegraphics[height=3.0in]{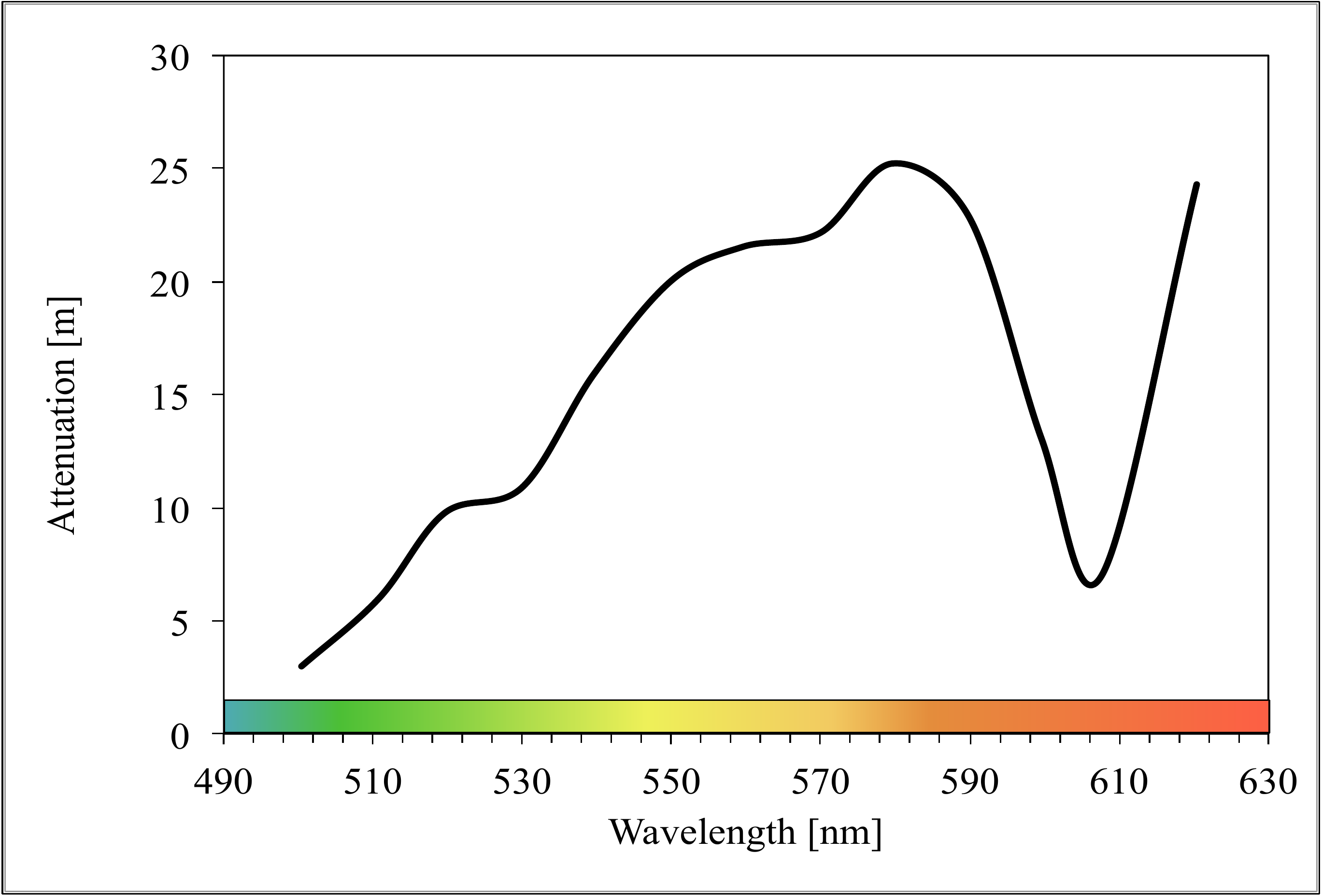}
\caption{Light attenuation length as a function of fiber emission
  wavelength.}
\label{fig:spectromatten}
\end{figure}

The spectrometer provides spectral measurements of the light emission
from the fiber. Model STS-VIS features a high spectral resolution
and high signal to noise ratio in the wide (350-800 nm) spectral
region. The spectral response from the fiber is obtained at several
points along the fiber. The result from a scan is shown on
Figure~\ref{fig:spectromreads} and suggests that the shorter
wavelength spectrum is attenuated at a significantly higher rate. The
spectrum at $\sim500$ nm is absorbed and re-emitted by the fluorescent
dye as this light component propagates along the fiber. We extract the
attenuation length values for various wavelengths
(Figure~\ref{fig:spectromatten}) by employing a single exponential fit
of the spectral light intensity as a function of distance to the
readout end.

\section{Conclusion}
We report the measurements of the light yield and attenuation of
Kuraray Y11 WLS fibers. We performed the using a
fiber scanner designed for the CRV system of the Mu2e experiment. The
fiber scanner features a fast and reliable data acquisition system,
and it will be an essential component for fiber quality assurance
during the production phase of the CRV. In addition, the fiber scanner
has been successfully used to study the properties of the fibers. The
results suggest a significant light yield gain from larger diameter
fibers. Kuraray Y11 WLS fibers yield long ($>10$ m) light attenuation
values for the 520 - 600 nm wavelength spectrum. Photo detectors with
high sensitivity in this spectral region are preferable for long
scintillating detectors with WLS fibers.

\Acknowledgments
I am grateful to Don Alfonso d'Alba for certain services essential to 
this investigation.

\end{document}

%% file: econfmacros.tex
\def\beq{\begin{equation}}
\def\eeq#1{\label{#1}\end{equation}}
\def\eeqn{\end{equation}}

\def\beqa{\begin{eqnarray}}
\def\eeqa#1{\label{#1}\end{eqnarray}}
\def\eeqan{\end{eqnarray}}

\let\bar=\overbar

\def\Dslash{\not{\hbox{\kern-4pt $D$}}}
\def\dslash{\not{\hbox{\kern-2pt $\del$}}}

\def\msb{{\bar{\ssstyle M \kern -1pt S}}}

